\documentclass[onecolumn]{mn2e}
\usepackage{graphicx}

\def\kms{\>{\rm km}\,{\rm s}^{-1}}

\def\Msun{\>{\rm M_{\odot}}}
\def\Ms{\Msun}

\def\tento#1{\times 10^{#1}}
\newcommand{\gtsim}{\mathrel{\hbox{\rlap{\lower.55ex \hbox {$\sim$}}
                   \kern-.3em \raise.4ex \hbox{$>$}}}}
\newcommand{\ltsim}{\mathrel{\hbox{\rlap{\lower.55ex \hbox {$\sim$}}
                   \kern-.3em \raise.4ex \hbox{$<$}}}}

\title{The Role of Primordial Kicks on Black Hole Merger Rates}

\author[Miroslav Micic, Tom Abel, Steinn Sigurdsson]
{Miroslav Micic$^1$\thanks{E-mail: micic@astro.psu.edu, tabel@slac.stanford.edu, 
steinn@astro.psu.edu}, Tom Abel$^2$ \& Steinn Sigurdsson$^1$ \\
$^1$ Department of Astronomy \& Astrophysics, Pennsylvania State University  \\
$^2$ Department of Physics, Stanford University  \\
}

\begin{document}
\maketitle

\begin{abstract}

Primordial stars are likely to be very massive $\geq30\Msun$, form
in isolation, and will likely leave black holes as remnants in the
centers of their host dark matter halos in the mass range
 $10^{6}-10^{10}\Ms$. Such early black holes, at redshifts z$\gtsim10$, 
could be the seed black holes for the many supermassive black holes found 
in galaxies in the local universe. If they exist, their mergers with
nearby supermassive black holes may be a prime signal for long
wavelength gravitational wave detectors.  We simulate formation of black
holes in the center of high redshift dark matter halos and explore 
implications of initial natal kick velocities conjectured by some formation
models. The central concentration of early black holes in present
day galaxies is reduced if they are born even with moderate kicks of
tens of km/s. The modest kicks allow the black holes to leave their
parent halo, which consequently leads to dynamical friction being less
effective on the lower mass black holes as compared to those still embedded 
in their parent halos. Therefore, merger rates may be reduced by more than 
an order of magnitude.  Using analytical and illustrative cosmological N--body
simulations we quantify the role of natal kicks of black holes formed
from massive metal free stars on their merger rates with supermassive
black holes in present day galaxies. Our results also apply to black
holes ejected by the gravitational slingshot mechanism.

\end{abstract}   


\section{MOTIVATION}

It is firmly established that most galaxies have super-massive
black holes (SMBH) at their centers. Their masses are in the range
10$^6 \, \Msun \ltsim M \ltsim 10^9 \, \Msun$ and it appears that there 
is a number of correlations (Kormendy $\&$ Gebhardt 2001, Gerhard 2001) 
between their masses and properties of the galactic bulge hosting them: 
mass of the SMBH on one hand and mass (Laor 2001) or luminosity or velocity 
dispersion (Merritt $\&$ Ferrarese 2001, Tremaine et al. 2002, Gebhardt et al. 2001) 
or light profile (Graham et al. 2001) of the galactic bulge on the other.
These correlations point to the link between the formation of SMBHs and the 
evolution of their hosts. It also appears that SMBHs are linked to the 
properties of the host dark matter halos. If the SMBH precursors have 
been present from very early on, then their mergers, together with growth
by accretion, could account for the abundance of the SMBHs 
today (Schneider et al. 2002).

Ab initio numerical simulations of the formation of the first luminous
objects in the current structure formation models find metal free stars 
to form in isolation and may have masses $30\,\Msun \leq m\leq 300 \,\Msun$  
(Abel et al. 2000, Abel et al. 2002, Bromm et al. 2002). In current models 
of structure formation, dark matter initially dominates and pregalactic 
objects form because of gravitational instability from small initial density 
perturbations. As they assemble via hierarchical merging, the metal-free 
primordial gas cools through rotational lines of hydrogen molecules. In the 
absence of metals, H$_2$ is the main coolant below $\sim$ 10$^4$K, which is 
the temperature range typically encountered in collapsing Population III 
objects. Rotational transitions of H$_{2}$, occurring via electric quadrupole 
radiation, allow the gas to cool and sink to the center of the dark matter 
potential well. This leads to a top-heavy initial stellar mass function 
and to the production of very massive stars, unlike the modern stellar IMF 
which is declining rapidly with increasing mass. 

Metal-free, high mass Population III stars die after 3 Myr, losing only a 
small fraction of their mass. If the fragmentation occurs during core 
collapse, two or more compact objects (black hole or neutron star) could 
be produced. This suggests that a large population of primordial massive 
black holes (MBH from now, with mass $\leq 10^3 \Msun$) could be an end 
product of such pregalactic star formation (Heger et al. 2003). Since they 
form in rare high-$\sigma$ density peaks (Couchman $\&$ Rees 1986, 
Madau $\&$ Rees 2001), relic MBHs would be predicted to cluster in the cores 
of more massive halos (Abel et al. 2002) formed by subsequent mergers. It has 
been suggested (Volonteri et al. 2003, Islam et al. 2003) that mergers of 
massive halos and clustering of these MBHs should start at redshifts as large 
as z$\sim$20. Growth of MBHs would proceed through accretion and coalescence 
which would lead to the formation of intermediate mass black holes (IMBH 
from now, with mass in range $10^3\,\Msun \leq m\leq 10^6 \,\Msun$). IMBHs 
can form sufficiently early on. In the most optimistic scenario (Haiman 
$\&$ Loeb 2001), 10$^9\Msun$ black hole could form at redshift z$\sim$10. 
However, through the mergers of dark matter halos at high redshift, black 
holes would form binaries and their coalescence under gravitational radiation 
would give them a significant kick velocity (Favata et al. 2004, 
Merritt et al. 2004). Anisotropic emission of gravitational waves which carry 
away linear momentum causes center of mass recoil. The recoil velocities are 
expected to be most likely in the range 10-100 kms$^{-1}$, kicks of a few 
hundred kms$^{-1}$ are not unexpected, and the largest recoils should not be 
above 500 kms$^{-1}$ (Favata et al. 2004). The amplitude of the kick determines 
if the black hole will be ejected from its host dark matter halo or if it 
will fall back. Even the most massive dark matter halo at redshift z$\geq$11 
can not retain the black hole that received 150$\kms$ kick while at the redhift 
of z$\geq$8, kicks of 300$\kms$ are sufficient to produce the same result 
(Merritt et al. 2004, Fig 3.). This is valid for black holes with masses  
$10^3\,\Msun \leq m\leq 10^8 \,\Msun$. Fully general relativistic simulations 
of radiation recoil from highly distorted Schwarzschild holes yield maximum 
kick velocities in excess of 400\,km/s (Brandt $\&$ Anninos 1999).  This 
should be sufficient to: eject IMBHs from globular clusters; displace IMBHs 
from the centers of dwarf galaxies; perhaps provide sufficient population
of IMBHs for merging at the centers of spiral galaxies (Merritt et al. 2004). 
In fact, for Kerr black holes, recoil is significant even for holes of equal 
mass (Favata et al. 2004) and may be directed out of the plane of the orbit 
(Redmount $\&$ Rees 1989). Kicks are also expected from the gravitational
slingshot - the ejection of one or more black holes when three black holes 
interact (Xu et al. 1994, Valtonen et al. 2000, Rees 1988). Space velocities 
much greater than those of their progenitores are common in neutron stars 
(Lai et al. 2001, Colpi $\&$ Wasserman 2002, Arzoumanian et al. 1997). Recent 
studies of pulsar proper motion implies characteristic velocities of the order 
of 200-500 kms$^{-1}$ (Wex et al. 2000), and as large as 100$\kms$ for black 
holes (Colpi $\&$ Wasserman 2002). The high space velocity of X-ray Nova Sco 1994 
implicates black hole kicks (Brandt et al. 1995).  Likely values of the velocity 
of the merged objects at infinity lie in the range of 100 - 300 $\kms$ 
(Davies et al. 2002).

Without exploring the nature of kicks, we are investigating the observational
implications of other peoples conjectures on black holes getting impulsive 
kicks, whether on formation or through mergers. We are also exploring the 
consequences that kicks will have on the IMBHs merger rates and the formation of 
SMBHs that are indirectly observed in the nuclei of nearby luminous galaxies. 
LISA will test the Population III IMBHs formation models. 

A powerful instrument for studying IMBHs will be LISA (Danzmann 2003). LISA 
can study much of the last year of inspiral, as well as the waves from the final 
collision and coalescence of IMBHS binaries, whenever the masses of the 
black holes are in the range $3\times 10^4 \, \Msun \leq M \leq 10^8 \, \Msun$. 
It can also study the final coalescences with remarkable signal to noise ratios:
S/N$\geq$1000. For black holes with the masses in the range 
100$\Msun\leq$M$\leq10^4\Msun$, out to cosmological distances, LISA can observe 
the last few years of inspiral, but not the final collisions. The equal-mass black 
hole binaries enter LISA's frequency band roughly 1000 years before their final 
coalescences, more or less independently of their masses, for the range
100$\Msun\leq$M$\leq10^6\Msun$. If the coalescence rate is one per year, LISA 
would see roughly 1000 additional IMBHs binaries that are slowly spiraling inward, 
with measurable inspiral rates. From the inspiral rates, the amplitudes of the two 
polarizations, and the waves' harmonic content, LISA can determine each such 
binary's luminosity distance, redshifted chirp mass (1+z)M$_c$, orbital 
inclination, and eccentricity. From the waves' modulation by LISA's orbital motion,
LISA can learn the direction to the binary with an accuracy of order
one degree (Thorne 1995, Cornish $\&$ Levin 2002, Vecchio et al. 2004).

In addition to the mapping of spacetime around SMBH with inspiraling
compact objects, BBO (Big Bang Observer) mission will have
arcsecond-arcminute precision in identifying and localizing every
merging black hole binary at any redshift, anywhere in the universe
over the years to months before their actual merger. In combination
with ground-based detectors at higher frequencies, it will measure the
mass, angular momenta, and dynamic spacetime structure of these compact
objects with unprecedented precision. 

If these IMBHs would accrete from the interstellar medium they may
also be found by deep X--ray observations (Agol $\&$ Kamionkowski 2002) in 
the Milky Way. This observational signature will depend sensitively on the spatial
distribution of these black holes in our galaxy. Similarly, the next
generation of micro-lensing searches may be able to constrain high mass
black holes from the longest events (Agol et al. 2002).

In our numerical simulations we use GADGET (GAlaxies with Dark matter 
and Gas intEracT), a code written by Volker Springel (Springel et al. 2001), 
for simulations of self-gravitating collisionless fluids evolution, with a 
tree method N-body approach, and a collisional gas using smoothed particle
hydrodynamics. 

In \S 2 we present straightforward analytical arguments. In \S 3 we describe 
the simulations setup, the codes used for analysis of the GADGET 
simulations, and present results for 3 different high resolution 
simulations with varying natal kick parameters for early
black holes.  In \S 4 we describe post-merger evolution through
dynamical friction, before we conclude in \S 5.

\section{ANALYTICAL EXPECTATIONS}

Clustering of MBHs at the centers of dark matter halos leads to their growth 
into the population of IMBHs at the centers of merging dark matter halos. 
In the following sections we will address every dark matter halo that hosts 
a black hole as a parent halo. The early stages of the merger are driven by 
the hierarchical cold collapse of the sub--halos (gravitationaly bound 
substructures of larger dark matter halos) into the primary halo (we focus 
our analysis on the single dark matter halo that naturally emerges from the 
numerical simulations as the largest in mass structure, hence the primary halo) 
forming the Galaxy.  Subsequent dynamical evolution of the IMBH population 
occurs through dynamical friction, and secular orbital evolution in the 
presence of any residual triaxiality after virialization (Madau $\&$ Rees 2001, 
Holley-Bockelmann et al. 2001).

We assume that IMBHs embedded in their parent sub--halos experience
dynamical friction acting collectively upon the entire compact sub--halo 
containing the IMBH, at least until tidal stripping significantly 
reduces the effective sub--halo mass to that of the central mass only 
(Weinberg 1989, Vine $\&$ Sigurdsson 1998).  Tidal dissolution becomes
effective only where the density of the primary halo is comparable to
that of the sub--halo. With characteristic length scales of ~$\sim
100 \,{\rm pc}$ and masses of $\sim 10^6\Msun$, sub--halo densities are
of order $10^{0 \pm 1}\Msun/pc^{3}$, and the primary halo density reaches 
such densities only in the inner few kpc, at radii $r \leq 0.01 \, r_{vir}$ 
(Fig 5).

The primary halo is well-approximated by a singular isothermal sphere
for radii of order $0.1\, r_{vir}$ (Madau $\&$ Rees 2001), the dispersion is near
constant and isotropic over a range in radii, and the density profile
is close to isothermal, with a steeper fall off at larger radii and
flattening at smaller radii.  











Approximating density interior to {\sl r}, the distance of the sub--halo of mass 
M from the center of the primary halo, with associated circular velocity v$_c$ to 
the center of the halo, timescale for dynamical friction to bring the sub--halo and 
associated IMBH to the primary halo center (Binney $\&$ Tremaine 1987, eq. 7.26) is:

\begin{equation}
{t_{fric}}=\frac{1.17}{ln{\Lambda}}\frac{r^{2}v_{c}}{GM}.
\end{equation}

where ln$\Lambda$ is Coulomb logarithm.

This is a conservative estimate for the time scale for early stages of
dynamical friction; the orbits of the sub--halos are not circular and
will therefore sink faster than we estimate, and coupling of the
dynamical friction to internal degrees of the sub--halo also
accelerates the evolution, possibly by as much as a factor of two
(Weinberg 1989, Vine $\&$ Sigurdsson 1998). At small radii, mass loss from tidal
disruption becomes significant, and the density profile flattens, but
the dispersion also decreases, and we may expect the IMBH to continue
their dynamical evolution towards the center of the primary halo, if
they were close enough to get inside ${r} \leq 0.01 r_{vir}$ in
the first place.

Typical mass ratios of early black holes and their host halos are 
$\geq10^{-4}$. Consequently for kicks that expel the IMBH from their 
host-halos at the higher redshifts, the dynamical friction time scale
is at least $10^{4}$ times larger. 

Assuming that dynamical friction is efficient in bringing sub--halo to 
the center of the primary halo, from the equation (1), we can calculate 
the radius $r_{sink}$ at which sub--halo has to be, in order to sink to 
the center in less than the Hubble time for given velocity dispersion and 
sub--halo mass. If every sub--halo carries one IMBH at its center, then 
the merger rate for IMBHs will be function of number of sub--halos inside 
$r_{sink}$. If kicks at higher redshifts supply IMBH with enough velocity  
for IMBH to escape its sub--halo, the expected number of IMBHs at 
${r} \leq r_{sink}$ will be smaller and this will lead to a significant
decrease in IMBH merger rates. Stopping the growth of IMBH through gas 
accretion by ejecting them into lower dark matter density regions changes
x-ray population predictions (Agol $\&$ Kamionkowski 2002). We illustrate 
these effects with N-body simulations in what follows.

\section{ILLUSTRATIVE SIMULATION}

\subsection{Simulation Setup}

We performed simulations that were set up for a cubic, periodic box of
14.3 comoving Mpc on a side in a $\Lambda$CDM universe with
$\Omega_M$=0.3, $\Omega_{\Lambda}$=0.7 and h=0.7 from redshift z=40 to
z=1. From initially low resolution simulations we selected a halo with
$2\tento{10}\Msun$ at redshift one. We then refined a sphere of 2 Mpc
comoving radius in the initial conditions and reran the simulations
from z=40. Using $128^3$ on the top level and a refinement factor of 4
in the high resolution region, we attain 4.9$\times10^6$
high-resolution particles (softening length 2 kpc comoving) for the 
simulation and 2.0$\times10^6$ low-resolution particles (softening 
length 4 kpc comoving) in the rest of the box. The mass of each high
resolution particle in these simulations is 8.85$\times10^5\Msun$
and the mass of each low-resolution particle is 5.66$\times10^7\Msun$.

\subsection{Analysis Tools}

Every snapshot consists of masses, positions, and velocities of all
types of particles at a specific redshift. Plots of their
positions show clumps of particles that need to be associated with
real-world objects. We used the HOP algorithm (Eisenstein $\&$ Hut 1998) to
divide the particles into distinct sets such that particles
in individual high-density regions are grouped together and left
separate from those in other regions. HOP assigns density estimates to
each particle by using the Gaussian smoothing algorithm and then
determines, of the particle and its nearest neighbors, which of the
particles has the highest density. Each particle is then associated
with its highest density neighbor and hopping is continued to higher
and higher densities until the highest density particle is
reached. All particles that hop to the same maximum are placed into
the same group. Every particle is assigned to one and only one
group. To distinguish between the dense halo and its surroundings, we
set a density threshold, cutting out of the group all particles with
densities less than 5$\%$ of the density of the densest particle. We
also define a halo as a group of more than 50 particles. The HOP
output is a list of halos with IDs, coordinates, density of the
densest particle, and number of particles in each halo. GADGET IDs of
the particles are used for following the particles from one snapshot
to another. They form different density distributions in different
snapshots and applying HOP analysis will give us different densest
particles in the same group in two different snapshots. HOP IDs are
just following the order in which the particles are stored in the
snapshot and they have to be matched with GADGET IDs if one wants to
follow the dynamics of halos in the simulations.

Further analysis is performed by Ganyl, Gadget Analysis Code (Abel et
al. 2001) and P-GroupFinder (Springel 2000). Ganyl analyzes spherical
profiles for gadget data, taking both the list of centers from HOP
analysis and a snapshot of the particle data that matches it. It
converts units back to proper from comoving and brings particles into
proper order according to their gadget IDs. Then, it identifies the
centers in the snapshot, and goes in spheres from them to find the
virial radii r$_{200}$, the total mass of the halos inside their
virial radii m$_{200}$, theoretical velocity dispersion, components of
angular momentum, and the total angular momentum. The virial radii are
defined as radii enclosing mass density two hundred times of the mean
mass density of the universe at the redshift of the snapshot.

P-GroupFinder is using a different approach for finding halos. Particle
distribution is segmented into groups using the friends-of-friends
(FOF) algorithm. Fraction of the mean interparticle separation is used
as linking length and halos are defined with distance criteria
only. Gravitationally self-bound substructures are then extracted from
each FOF group by using SUBFIND algorithm. The bound part of FOF halo
counts as subhalo as well, so every FOF halo has its bound
counterpart.

These two different approaches in identifying halos are
compared. Advantage of Ganyl is in its cosmological approach of
defining halo using over-density, but since Ganyl is using density
centers from the HOP algorithm, it is finding over-densities more than
once for the same object creating artificial halos in this way. Since
P-GroupFinder is checking for gravitational boundness, it can isolate
one of these artificial halos as a real world object. Virial radii in kpc 
are calculated from total mass in particles (100$^{1/3}\times{M_{tot}}^{1/3}$) 
and their values are slightly larger than virial radii calculated by Ganyl. 
We use P-GroupFinder to identify black holes as the most bounded particles 
in their host halos.

We construct a merger tree from all 33 snapshots. The most massive
halo incorporates hundreds of subhalos inside its virial
radius. We examined all mergers and selected a sample of the most
massive halos from each group of mergers together with a sample of
isolated halos that do not merge. For this new list we found GADGET
IDs of the centers of every halo and by tracing IDs in different snapshots, 
we identified coordinates and velocities to get trajectories of
halo centers.

Density plots were made with codes provided by Naoki Yoshida which
apply a spline SPH kernel to derived densities and then project them
along one axis to produce a two dimensional smoothed image.

\subsection{Black Holes Trajectories}

 The most massive halo in our simulation, at redshift z=1, (we will
refer to this halo as primary from now on) has virial radius
R$_{vir}$=370 kpc proper, and its mass inside this radius is
1.52$\times10^{12}\Msun$.  We focused on studying properties of the 
primary halo and its substructure.

We assumed that at redshifts higher then z=8, every halo has an IMBH at 
its center. As Pop III remnants, formed before most other structure, 
IMBHs would have enough time to settle inside gravitational potentials 
of dark matter halos. From the simulation point of view this population 
can be selected by choosing every most-bounded particle in the halos 
discovered by P-GroupFinder. Tracking of this population through time 
can tell us how many IMBHs we can expect today inside galaxies and
what their distribution would be. At redshift z=8.16 we identified
2869 dark matter halos with mass in range $10^7 \Msun \ltsim M \ltsim
10^{10} \, \Msun$. We selected the same number of IMBHs from their
centers. By connecting particles' coordinates through 33 snapshots, we
obtained MBHs trajectories from redshift z=8.16 to redshift z=1.00. 
From 2869 IMBHs at z=8.16, 1958 of them can be found inside primary halo 
at z=1.00. Fig 1. (top) shows density plot of XY-projection of SIM1 box 
at redshift z=8.16. Density peaks (in yellow) are the centers of dark 
matter halos selected by P-GroupFinder and in the same time the positions 
of particles which we selected for IMBHs. A sample of their trajectories 
(trajectories of IMBHs from the hundred most massive halos) as they spiral 
inside the primary halo is overploted. It is suggested 
(Miralda-Escude $\&$ Gould 2000) that clusters of IMBHs might exist 
inside the inner kpc of every galaxy, spiraling toward SMBH at the center.

\subsection{Black Hole Kick Velocity}  

As mentioned in the introduction, coalescence under gravitational 
radiation may give an IMBH a significant kick velocity.  We model 
the distribution of natal kicks with a truncated Gaussian in the 
interval $\{0,150\} \ {\rm km \, s^{-1}}$, with a mean kick of 
$75\ {\rm km \, s^{-1}}$.  By an ``IMBH'' we are referring to the 
2869 particles which define the centers of those sub--halos, at 
redshift 8.16, 3/4 of which are identified as destined to eventually 
merge with the primary halo. We add this presumed kick velocity with 
a random direction to a gadget velocity taken from the snapshot at 
redshift 8.16, for every particle identified as tracing the location 
of a presumed pop III IMBH. In this way, we obtain a new snapshot 
file with changed velocities of the group of black holes only. We use 
this snapshot file as an initial conditions file for a new simulation 
(SIM2a from now on) that starts at redshift 8.16, and we examine the 
differences in trajectories of the particles identified as tracers of 
the black holes, with and without natal kick velocities. The same 
procedure was repeated for different interval of velocities 
$\{125,275\} \ {\rm km \, s^{-1}}$, with a mean kick of 
$200\ {\rm km \, s^{-1}}$ which was used for simulation - SIM2b. 
These are the most likely ranges suggested in the related works 
(Favata et al. 2004, Merritt et al. 2004).

Notice the fundamental difference between SIM1 and SIM2. In the case 
without the kicks, black holes embedded in sub--halos reach the center 
of the primary halo and the location of the presumed SMBH through 
dynamical friction. The main contribution to this procces comes from 
the total mass of their halos which remain mostly bound through our 
simulation.  In SIM2, IMBHs are generally ejected from their sub--halos 
by assigning new velocities and dynamical friction is not expected to be 
as efficient as in the first case since the dynamical friction mostly 
acts upon the bound sub--halo, not the much lower mass black hole 
(Hansen $\&$ Milosavljevic 2003, Gerhard 2001, Portegies Zwart 2003). 
Fig 1. (middle) and Fig 1. (bottom) show the density plot of XY-projection 
of SIM2a box (middle) and SIM2b box (bottom) at redshift z=8.16. As in 
SIM1, trajectories (from z=8.16 to z=1) of a sample of IMBHs are over-plotted. 
The same population of IMBHs, in three scenarios which differ only in
IMBHs' velocities, will have different trajectories depending on the kick
assigned. In the first place, kicks have to be large enough to provide
IMBHs with a velocity larger than the host halo escape velocity. If not, 
IMBH will not be able to escape the gravitational potential of the host 
halo and this will lead to a scenario quite similar to SIM1 (compare top 
and middle Fig 1.). Larger kicks, as in SIM2b, change IMBHs' trajectories 
dramatically (compare top and bottom Fig 1.). Fig 2. explains the 
difference. Marked as pluses are the maximum escape velocities calculated 
from the gravitational potential of every dark matter halo selected at 
z=8.16. Escape velocity decreases with halo radius and reaches its maximum 
at the center. Hence,  the maximum escape velocity is the escape velocity 
of IMBH since IMBH is the particle at the halo center. Escape velocity 
increases with the halo mass. Shown as circles are the velocities of IMBHs 
relative to their host halos at z=8.16 after assigning them kick centered 
at 75 $\kms$. The plot shows that IMBHs from less massive host halos have 
velocities larger than escape velocities. With the increase of host halo 
mass, IMBHs need to acquire larger velocities in order to escape from host 
halo gravitational potential and, for some of them, assigned kicks are not 
large enough (all the IMBHs that lie below data set are represented with 
pluses). These IMBHs continue the host halo original path toward primary 
halo as in the case of no kick. This process is demonstrated by the IMBHs 
in SIM2a (Fig 2. circles), hosted by most massive halos (kick velocities 
less than escape velocities). Since we ploted trajectories in Fig 1. only 
for IMBHs hosted by the hundred most massive halos, trajectories in 
Fig 1.-top and Fig 1.-middle are almost identical). This same set of IMBHs, 
but now with kick centered at 200 $\kms$ in SIM2b (diamonds), almost all 
have enough velocity to escape host halos. As a result, trajectories in 
Fig 1.-top substantialy differ from the trajectories in Fig 1.-bottom. 
Unlike Fig 1.-top and Fig 1.-middle, the IMBHs in Fig 1.-bottom leave their 
host halos at z=8.16 and as a result do not form binaries on their path to 
the primary.  

Even though a substantial kick is assigned to black holes, and the dynamical
friction does not play an important role for them anymore, they still manage 
to merge with the primary halo and some fraction of them makes way to the 
center to coalesce with the SMBH at the center of the halo. As the 
gravitational potential of dark matter halos increases at lower redshifts 
(Fig 3), kicked IMBHs are being captured and sink to the primary. Fig 3. 
shows maximum escape velocity as a function of dark matter halos' mass at 
different redshifts for all halos identified in our simulation. The primary 
halo (thick line) goes through a large increase in gravitational potential 
toward lower redshifts. The maximum escape velocity from the primary halo 
increases from $\sim250\kms$ at z=8.16 to $\sim700\kms$ at z=1. Hence, the 
growth of the primary halo and the deepening of its gravitational potential 
well is responsible for the capture of IMBHs originally ejected from their 
host halos. However, the final spatial distribution of the presumed IMBHs 
is quite distinct (Fig 4), and their subsequent rate of coalescence with the 
central SMBH consequently may be different by orders of magnitude. 

The number of IMBHs inside the primary halo at z=1 is 1958 and in SIM2a
the number is only 14 IMBHs smaller. This is a coincidence, caused by the 
fact that we have changed velocities of all IMBHs identified at z=8.16 
and not just of those that are being identified inside the primary halo at 
z=1. This means that some number of IMBHs that did not reach the primary 
halo originally, reached it in this new simulation because of the changed 
velocities and hence directions. Clearly the new IMBH arrive ``naked'', 
without being enveloped in the dark matter sub--halos in which they formed 
originally. The number of IMBHs inside the primary halo in both SIM1 and 
SIM2a originating from the same sub--halos, is 1851.  Thus, 5.4$\%$ of the 
original population found in the primary halo, after change in their 
velocities at z=8.16, will not be found there. Similarly, a comparable number 
of the IMBHs that originally did reach the primary halo, will not in the 
repeated simulation.

Changes in trajectories of IMBHs in the repeated simulation with larger
kicks (SIM2b) is more pronounced (Fig 1. bottom). With larger kicks, more
IMBHs attain velocities large enough to leave their host halos. Although 
they need more time to sink into the potential well of the primary halo, 
1795 of them reach the primary halo at redshift z=1. The number of IMBHs 
originating from the same sub--halos as in SIM1 is 1630. Thus, 16.7$\%$ 
of the original population is lost due to the kicks, but for the same 
reason, a new population of IMBHs is kicked into the primary halo, which 
in total gives 1795 IMBH - meaning that the primary halo in SIM2b will 
have only 8.3$\%$ of IMBHs less than the primary halo in SIM1.

\section{Post-merger evolution}

We find that significantly more black holes get at the center of the
primary halo when they are embedded in their dark matter sub--halos.
But a number of mergers to the center occur even in the presence of
kicks.  Fig 4. shows the number density of the IMBHs inside the primary 
halo, from SIM1, SIM2a and SIM2b. Although the total number of black holes 
differs from SIM1 by 0.7$\%$ for SIM2a and 8.3$\%$ for SIM2b, there is 
a decrease in the number of IMBHs at SIM2 for the inner 10$\%$ of the 
primary halo.

We now estimate the dynamical friction time scale for the IMBH and 
sub--halo sinking in the gravitational potential of the halo. From
eqn 1), for the singular isothermal sphere with circular velocity 
v$_c$, the velocity dispersion is $\sigma$=v$_c$/$\sqrt{2}$, the 
virial radius of the halo is r=r$_{vir}$, and the mass inside this 
radius is 1.52$\times10^{12}\Msun$. For the primary halo these values 
are r$_{vir}$=370kpc comoving, velocity dispersion $\sigma=157\ 
{\rm km\, s^{-1}}$, and impact parameter b$_{max}$=r$_{vir}$. We 
calculate a radius ($r_{sink}$) which IMBHs have to reach in order 
to merge to the center in less than the Hubble time. We distinguish 
two cases. First, when the IMBHs are inside sub--halos of minimum 
mass 10$^7\Msun$ (SIM1). Second, when IMBH has been ejected from its 
parent sub--halo (SIM2).  In the first case, IMBH is brought to some 
radius inside the primary halo by the parent sub--halo and will 
continue sinking toward the center embedded within it. If we plug in 
Hubble time in formula (1), for the above values of velocity dispersion 
and mass we calculate that due to dynamical friction IMBHs inside 
10$^7\Msun$ will merge to the center if they are at less then 
$r_{sink}$=r$_{vir}/30$ when the halo collapse virialises. We find that 
a little over 4$\%$ of the IMBHs formed are at radii less then $r_{sink}$. 
So for this model we predict that in the absence of kick, 83 IMBH reach 
the center to coalesce with the central SMBH (or the seed SMBH formed in 
the sub--halo that became the center of the primary halo).  More generally, 
this predicts ${\it O}(10^2)$ IMBH mergers per Milky Way like halo over 
a Hubble time, even for the cases shown here where only halos with masses 
$\geq2.83\times10^{7}$ were allowed to form black holes. Since there are 
$\sim$10$^{10}$ galaxies in the observable universe in this mass range,
the LISA' merger rates will be $\sim$10$^{12}$ per Hubble time or R$\sim$100$/year$. 

In the second case, the kicked IMBHs have a significantly flatter spatial 
distribution, partly because they have decoupled from their parent 
sub--halos, so there are fewer inside $1/30 r_{vir}$. Fig 4. shows that 
there are 2.21$\%$ of IMBHs from SIM2a and 0.95$\%$ of IMBHs from SIM2b 
inside this radius. Fig 5. shows the ratio of IMBHs with kicks and IMBHs 
with no kicks. There is a large drop in the central population of IMBHs 
with kicks. This can be seen also in density plots Fig 6. By repeating the 
same calculation as in the previous case with the only difference in the 
value for mass in the equation for dynamical friction, 8.85$\times10^5\Msun$ 
instead of $10^7 \Msun$, (notice that we use particles as tracers of IMBH, 
in reality IMBH has smaller mass and needs to get to smaller radii for 
dynamical friction to be efficient) since that is the mass of single particle, 
we calculate that in order to merge at the center, IMBHs in SIM2 have to be 
at radii less then $r_{sink}$=$r_{vir}/100$.  Only 1/4$\%$ of the SIM2a IMBHs 
are inside this radius. That is, under the same assumptions but allowing for 
natal kicks, only about 4 to 5 IMBH merge with the central SMBH over Hubble
time, a factor of 20 lower merger rate. In the SIM2b there are no IMBHs inside 
$r_{vir}/100$ except for the one originating from the ancestor of primary halo 
at redshift z=8.16. These numbers can increase since, in reality, $\sigma$ 
is lower at small radii so $t_{fric}$ is also smaller. The merger of 83 IMBHs 
in SIM1 leads to formation of 7$\times10^7\Msun$ SMBH inside the dark matter 
halo with velocity dispersion $\sigma=157\ {\rm km\, s^{-1}}$. From the 
cosmological Monte Carlo realizations of the merger hierarchy 
(Volonteri et al. 2003), SMBH of the same mass and halo velocity dispersion 
would be formed only when mergers are combined with gas accretion as mechanisms 
for SMBH formation. Both results are below empirical values (Ferrarese et al. 2002). 
We note that the primary halo is not actually spherical and if a halo is triaxial, 
then some fraction of the IMBHs can ``walk'' into the inner halo 
(${r} \ll 10^{-2} r_{vir}$) region on time scale $\sim10 t_{orbital}$ due to 
centrephilic box or boxlet orbits (de Zeeuw 1985, Zhao et al 2002, 
Holley-Bockelmann et al. 2005), at which point dynamical friction becomes
effective in bringing the IMBH to the halo center.  Also, some of the
IMBHs are being assigned with kick velocities that directed them toward
center. Due to this, some additional IMBHs from SIM1 and SIM2 could
reach orbits in the center of HALO 1. We will explore the consequences
of the non--spherical shape of the primary halo on the long term
dynamical evolution of the sub--halos, and the implications for black
hole mergers and evolution of the inner regions of the galaxy in
another paper (in preparation).

Kicks are also responsible for ejecting IMBHs from the gas enriched regions
of the halos. Since gas accretion is one of the main mechanisms for black
hole growth, dumping black holes into regions of lower gas densities would 
prevent formation of AGNs which would lead to a decrease in their numbers.
Gas accumulates where the gravitational potential wells are deepest. It is
also where dark matter densities are highest. Therefore we may get a crude
estimate of the ability of IMBHs to accrete gas by tracing the dark matter 
density at their positions. Fig 7. shows average local density of dark matter 
traced by our set of IMBHs vs. redshifts. At z=8.16 the kicks are assigned. 
IMBHs are being ejected from their host halos into the regions with lower 
density (dash and dash-dot from z=8.16 to z=7.75). From z=7.75 to z=2.80 
kicked IMBHs are in the environment 1-6 times less dense then in the no-kick 
case. This might suppress formation of AGNs. Notice that around z=3.00, the 
average IMBHs in all three cases start tracing similar density distributions. 
This implies that for z$\geq$3, the contribution of faint AGNs to the ionizing 
background would be decreased if kicks of IMBHs were important.

\section{DISCUSSION}

We have performed and analyzed high-resolution collisionless simulations 
of the evolution of structure in a $\Lambda$CDM model. We have followed 
the formation and evolution of dark matter halos and by assuming that 
each halo is a host of one IMBH we have studied the formation of SMBH. We 
focused on two specific cases. First, IMBHs together with their host 
halos merge through dynamical friction. Second, when IMBHs are endowed with 
an initial kick, this leads to the ejection from their host halos in many 
cases. Analytically it is clear that the dynamical friction will act more 
efficiently on the host halos than on the much lower mass black holes 
formed within them. Our illustrative calculations highlight some of the 
expected differences in the density distribution of the final distributions 
of black holes which may be quite different, even in the presence of the
modest kick velocities we have imposed.

In order for dynamical friction to work in numerical simulations, the 
density of the background - primary halo density - has to be more 
resolved. Larger resolution gives smaller softening lengths for 
particles and smaller softening lengths give more realistic dynamical 
friction. At the current resolution of our simulation, dynamical 
friction can not be efficiently realized in the model after z$\sim$1 
and r$\ltsim$r$_{vir}/30$ because the background density of parent 
halos is not resolved well enough. Even at the current resolution, our 
simulations are able to account for effects of dynamical friction on 
the sub-halos at high z. Subsequent papers will track the dynamical 
evolution of the sub-halos and IMBH at late times, using higher 
resolution simulations and semi-analytic implementations of dynamical 
friction (Binney $\&$ Tremaine 1987). A little over 4$\%$ of IMBHs 
merge at the center in less than Hubble time. If this is the dominant 
way of creating SMBH then it is efficient even if the IMBHs are ejected 
from their parent halos, with merger numbers only reduced by a factor 
of two or so for modest kicks. It is also possible that kick velocities 
have been underestimated in our simulations. The value of the median 
kick of 150km/s may be increased to more then 1000km/s according to 
some authors (Favata et al. 2004, Merritt et al. 2004) which would also 
greatly enhance the effects introduced here, reducing the number of 
mergers too negligibly small. LISA observations should strongly constrain 
any natal kick on IMBHs formed from very massive Pop III stars in low 
mass proto-galactic sub-halos.



\clearpage

\begin{figure}
\begin{center}

\includegraphics [width=2.5in,angle=270]{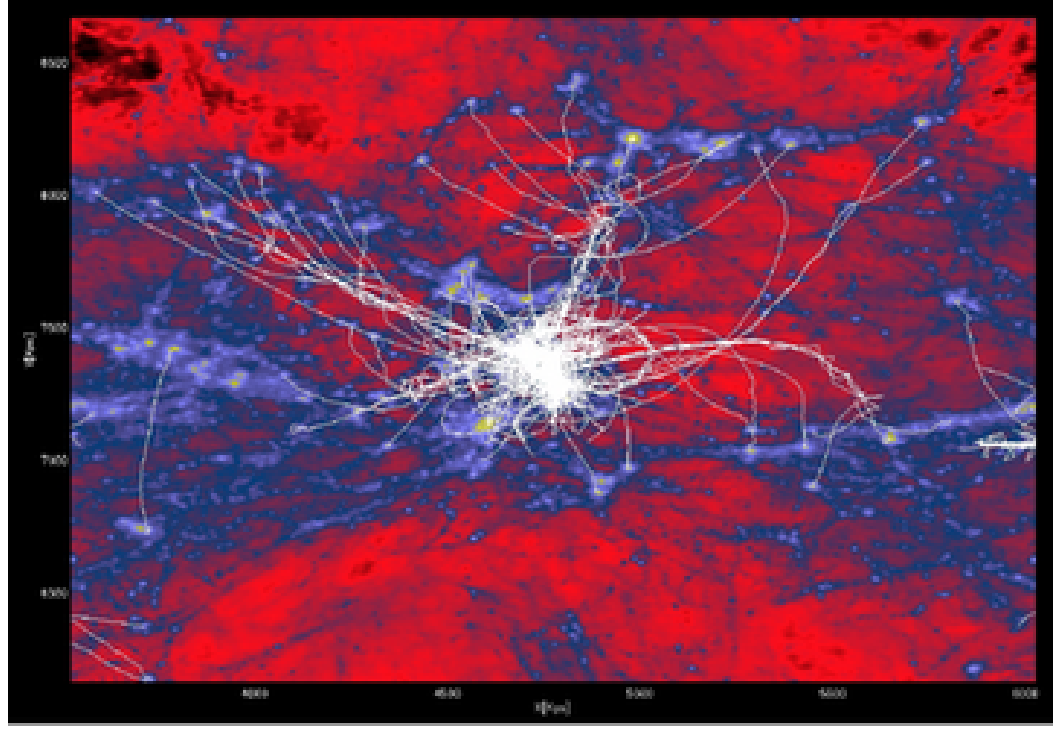}
\newline
\includegraphics [width=2.5in,angle=270]{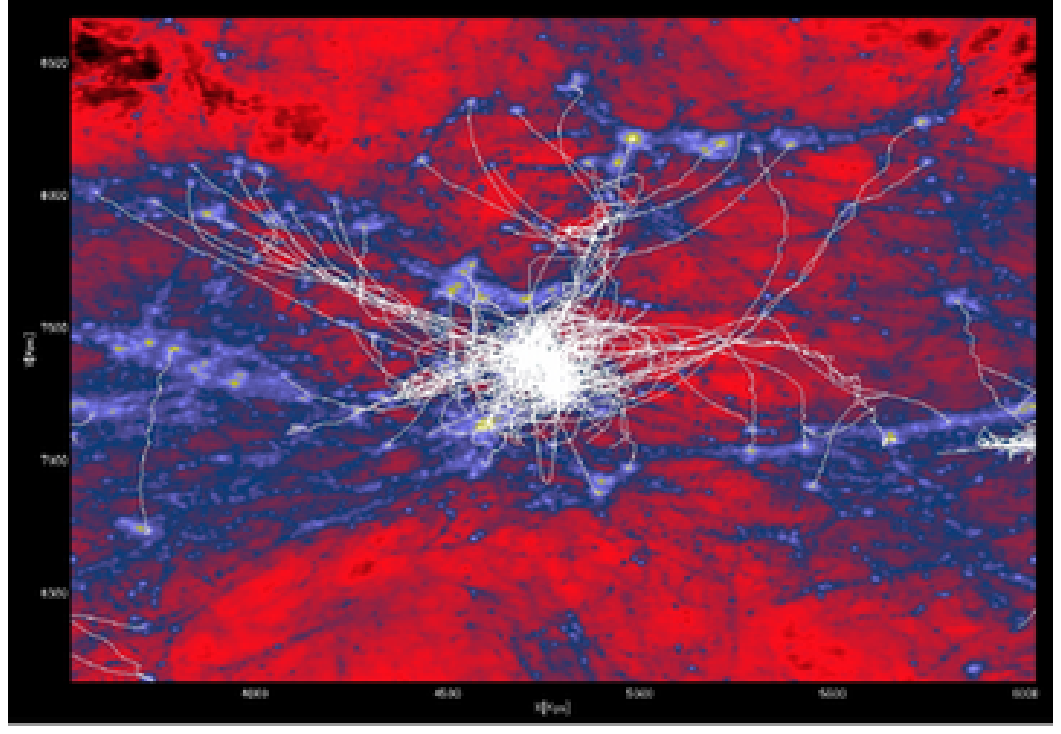}
\newline
\includegraphics [width=2.5in,angle=270]{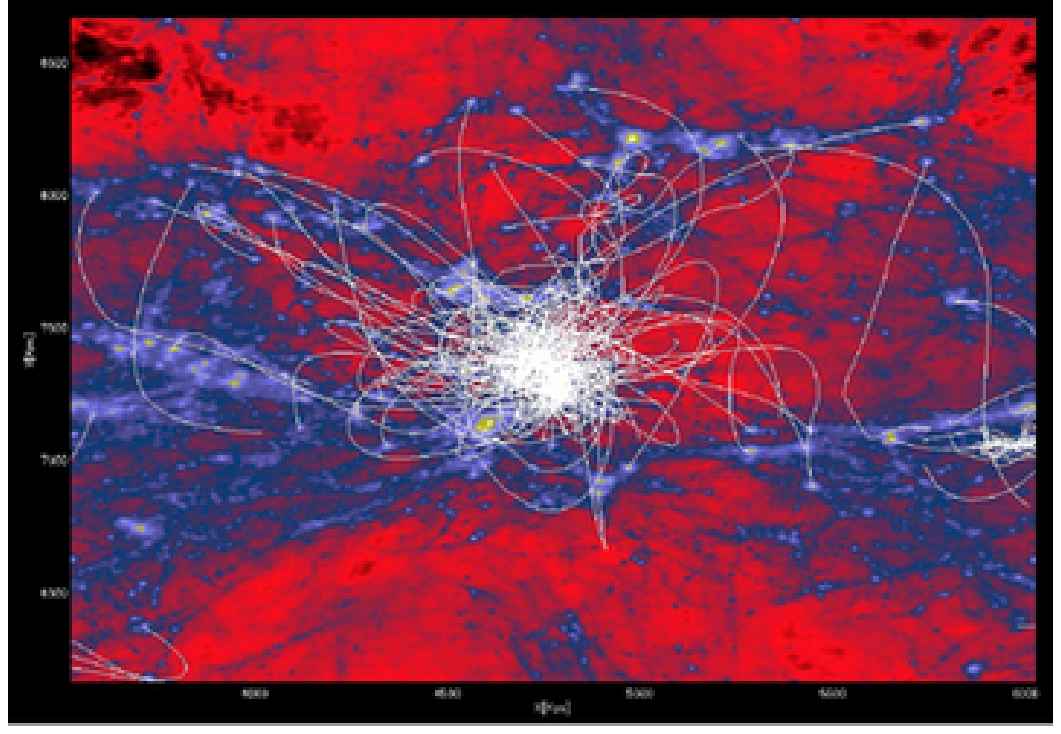}
\newline
\caption[Fig 1.]{Sample of hundred IMBHs selected from most massive host 
halos at z=8.16 and their trajectories (white) from z=8.16 to z=1 overploted 
on 2D density projection. Density peaks in yellow correspond to host
halos centers at z=8.16 and to the positions of their IMBHs. top: no kick 
case (SIM1); middle (SIM2a): case of [0,150] km/s kick centered at 75km/s and 
bottom (SIM2b): case of [125,275] km/s kick centered at 200km/s. With large 
kicks IMBHs overcome the host halos gravitational potential resulting in change 
of their trajectories and the final distribution of IMBHs (smearing of the 
trajectories at the center, quantified in the later plots). Observe, e.g., the 
upper right corner of all tree plots where two IMBHs form a binary in SIM1 and 
SiM2a. In SIM2a, IMBHs are trying to escape but the gravitational potential of 
dark matter halos is pulling them back. In SIM2b, assigned kick is larger then in 
SIM2a and also large enough to prevent formation of binary. Underneath this binary 
in SIM1, at the coordinates [5700,7000]kpc, a dense halo center hosts an IMBHs. In 
less then 100kpc, this halo captures three IMBHs more. This small ``cluster'' of 
IMBHs spirals in toward the center. The picture clearly resolves IMBHs orbits around 
center of mass up to the point, [5200,7400]kpc, where the orbital separation is smaller 
then the softening lenght. This point is reached later in SIM2a because of the kicks 
while in SIM2b binaries never form.}

\end{center}
\end{figure}

\clearpage


\begin{figure}
\vspace{0.5in}
\begin{center}
\includegraphics [width=4.5in,angle=0]{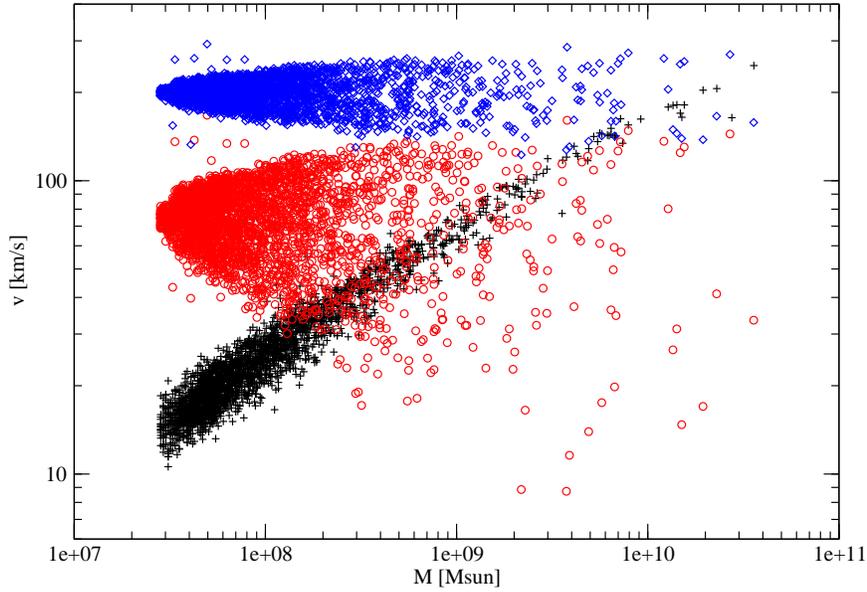}
\caption[Fig 2.]{Black holes' escape velocity as a function of halo mass 
(z=8.16) as pluses, calculated from gravitational potential of host halo set 
consisting of 2869 members. Black holes' velocity relative to their host halos' 
velocity reduce to assigned kicks: [0,150] km/s centered at 75 km/s repesented 
as circles and [125,275] km/s centered at 200 km/s represented as diamonds. 
Lower mass host halos have lower maximum escape velocities enabling their IMBHs
to escape. IMBHs in more massive host halos demand larger kicks in order 
to escape. As a result, some of them stay captured in their host halo 
gravitational potential.}

\end{center}
\end{figure}



\begin{figure}
\vspace{0.5in}
\begin{center}
\includegraphics [width=4.0in]{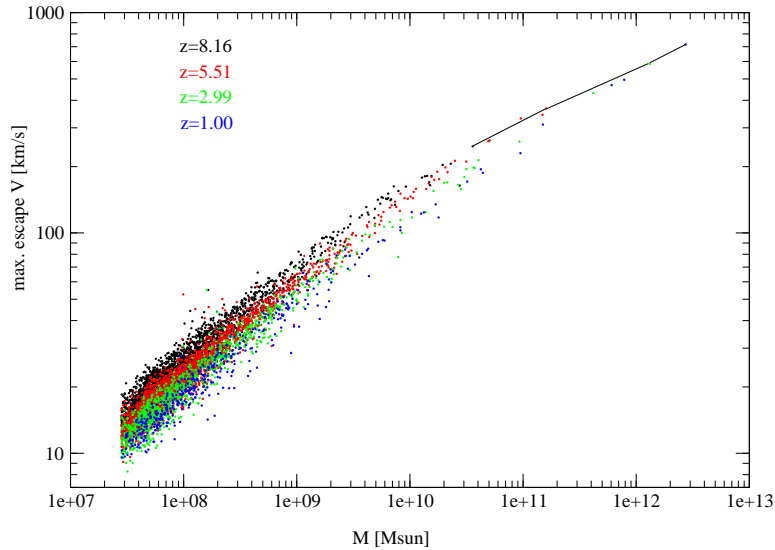}
\caption[Fig 3.]{Maximum escape velocity (corresponds to central gravitational
potential) of dark matter halos at different redshifts as a function of dark 
matter halos' masses. Gravitational potential of the primary halo (thick line) increases 
at lower redshifts capturing even the IMBHs which have been assigned with the 
highest kick velocities.}

\end{center}
\end{figure}

\clearpage

\begin{figure}
\vspace{0.5in}
\begin{center}
\includegraphics [width=3.0in,angle=270]{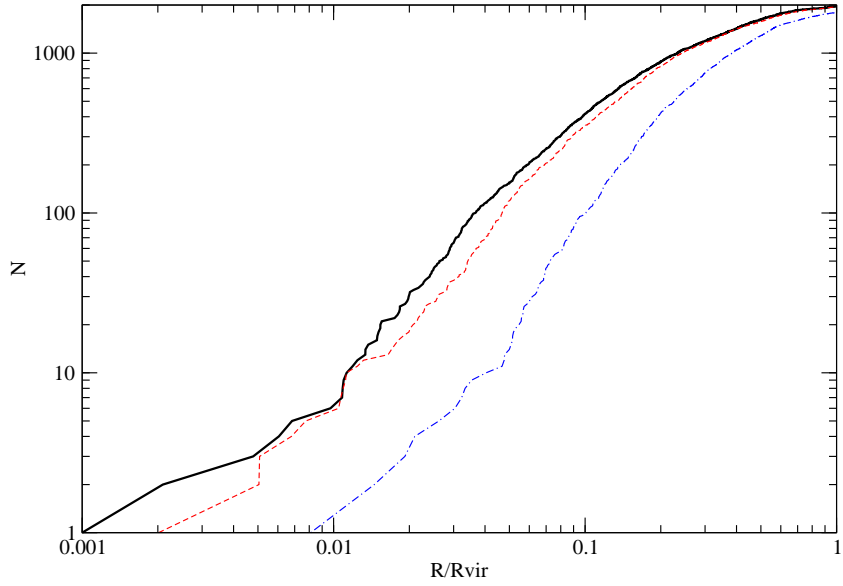}
\caption[Fig 4.]{Number of black holes as a function of primary halo radius. 
No kick case (thick); [0,150] km/s kick centered at 75 km/s (dash) and 
[125,275] km/s kick centered at 200 km/s (dash-dot). Although the number of 
IMBHs entering primary halo is comparable in all three cases, the difference
in their interior distribution is well pronounced.}
\end{center}
\end{figure}

\begin{figure}
\vspace{0.5in}
\begin{center}
\includegraphics [width=3.0in,angle=270]{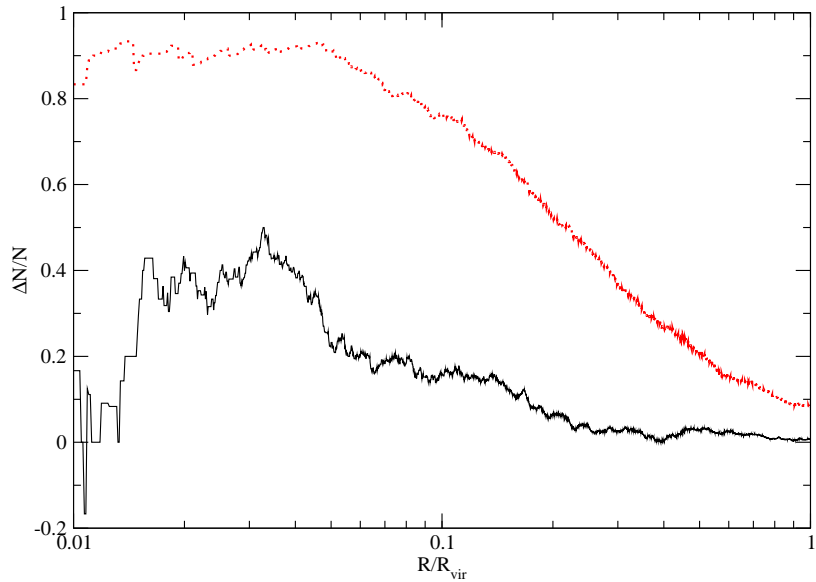}
\caption[Fig 5.]{Fraction of SIM1 black holes in SIM2a and SIM2b as a 
function of radius. thick line for [0,150] km/s kicks centered at 75 km/s and 
dots for [125,275] km/s kicks centered at 200 km/s.}
\end{center}
\end{figure}

\clearpage

\begin{figure}
\vspace{0.5in}
\begin{center}
\includegraphics [width=3.0in,angle=270]{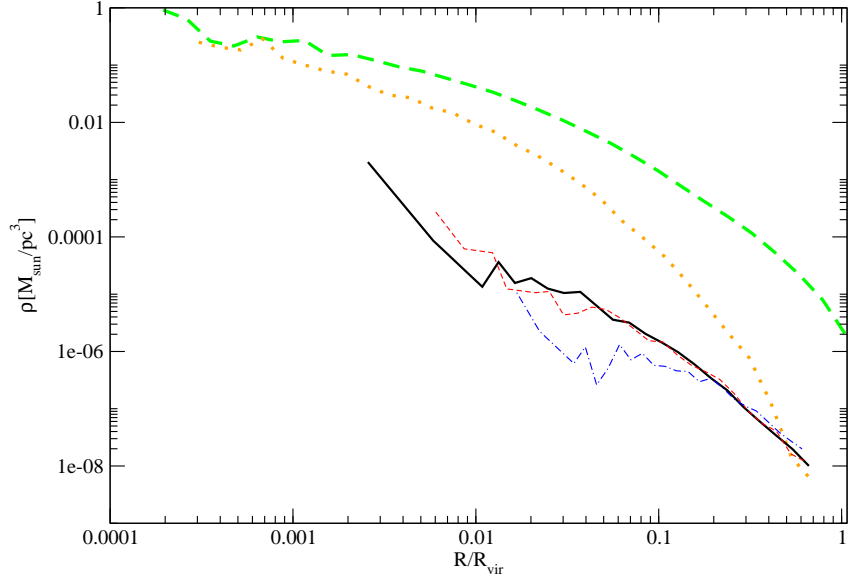}
\caption[Fig 6.]{Density of the primary halo at z=1 (thick dash) and its most 
massive progenitor from z=8.16 (dots) as a function of radius. Also, density 
in hosted black holes for no kick (thick); [0,150] km/s kick centered at 
75 km/s (dash) and [125,275] km/s kick centered at 200 km/s (dash-dot).}
\end{center}
\end{figure}

\begin{figure}
\vspace{0.5in}
\begin{center}
\includegraphics [width=3.0in,angle=270]{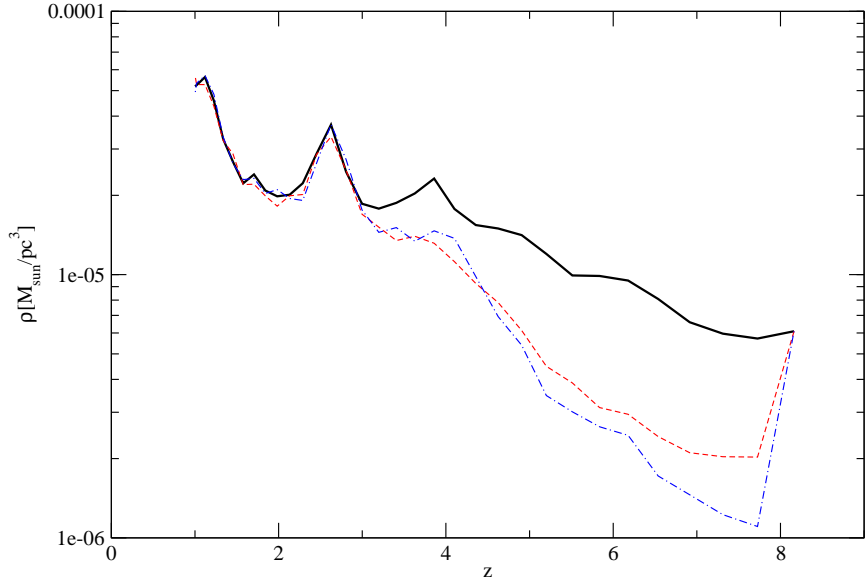}
\caption[Fig 7.]{Local density of dark matter traced by black holes as a 
function of redshift. No kick (thick); [0,150] km/s kick centered at 75 
km/s (dash) and [125,275] km/s kick centered at 200 km/s (dash-dot). Ejection 
of IMBHs from gas enriched regions of galaxy infuences AGNs formation 
rates, reduces their numbers and their contribution to ionizing background.}
\end{center}
\end{figure}

\end{document}